\documentclass[journal]{IEEEtran}   

\IEEEoverridecommandlockouts                              

\usepackage{graphicx}
\usepackage{epstopdf}
\usepackage{amsfonts}
\usepackage{amsmath,amssymb}
\usepackage{color}
\usepackage{fancyheadings}
\usepackage{tikz}
\usetikzlibrary{automata}
\usepackage{color}
\usepackage{verbatim}
\usepackage{algorithm}
\usepackage{subcaption}
\usepackage{booktabs}
\usepackage{tabularx}
\usepackage{rotating}
\usepackage{caption,subcaption}
\usepackage[switch]{lineno}

\usepackage[margin=0.65in,footskip=0.25in]{geometry}

\begin{document}
\title{\LARGE \bf Shape the Future of ITS -- \\
\large Optimization and scheduling for a large scale urban transportation system --\\
in a fast-changing world}

\author{Yi Zhang
\thanks{Yi Zhang is affiliated with with the Institute for Infocomm Research (I2R), Agency for Science, Technology and Research (A*STAR), Singapore 138632. Emails: yzhang120@e.ntu.edu.sg(zhang$\textunderscore$yi@i2r.a-star.edu.sg).}
}

\maketitle

\section{Introduction}
The rapid growth of the population facilitates the urban sprawl, automobile production and leads to heavy traffic congestions, pollution, noises and traffic fatalities. Therefore, it is indispensable to develop an eco-friendly sustainable transport with high performance of the existing traffic network. Although many transformational products have been developed and used on the road, such as autonomous vehicles (AV) and drones, and their quality is continually upgraded, 100\% replacement of traditional cars to AVs requires not only the safety and security evaluation but also corresponding physical infrastructures, e.g., non-signalized intersection, charging stations. Also, we still have a long way to go in order to completely commercialize AVs with level 5 automation. Thus, it is predictable that traffic signals will still play an important role for a certain long time in this hybrid transportation world, which involves both human-driven cars and AVs under different automated levels. At the period of transition, a feasible and implementable approach is more helpful and practical to reshape the transportation system in the near future.

The concept of the pedestrian/transit-oriented transport, which is designed for the human body instead of the car body, has been proposed by researchers \cite{IET}: ``cities with very high walking, cycling and transit mode share (i.e., 75\% or more) typically have high density, mixed use urban centres at or above 100-200 people per hectare and are supported by a transportation strategy that prioritizes pedestrians first, then cyclists followed by transit users." However, vehicle traffic still gets excessively high attention in many cities' guidelines, and walkability is seldom taken into account. Also, walking is the access to public transport (PT), thereby PT services could accordingly be promoted if the pedestrian walking safety and pleasantness can be satisfied. PT system has been studied for several decades due to its large ridership
and sustainability on economic efficiency, environmental protection and social equity. Proper bus dispatching and operation can attract more passengers, which encourages commuters to change their travel mode from private automobiles to public buses to further alleviate the traffic congestion and air pollution.

However, PT system, a typical ride-sharing transport serving a variety of access needs and providing an equal social value, now faces the challenges from the mobility-on-demand (MOD) system, which is famed for its easy transactions and convenient access via mobile phones. Therefore, we need to explore and create a win-win cooperation model between both PT and AV systems at current transitional period. The flexible demand-driven pattern in AVs could make up the shortages due to fixed routes in the PT system. On the other hand, PT system is still coping with the high-volume transfer tasks, which may not be feasible for AVs due to the safety concern.

The objective of this essay is to propose a set of coordinated technological solutions to transform existing transport system to a more intelligent interactive system by adopting optimization and control methods implementable in the near future, thereby improving public services and quality of life for residents. In this essay, three different application scenes that closely related to people’s daily life are discussed.
We firstly propose a traffic light scheduling strategy via model predictive control (MPC) method, with the aim to fairly minimize both pedestrians' and vehicles' delay. After that, a combined dispatching-operation system is proposed to further increase the control flexibility, and corresponding implementation solution for boarding control is also illustrated. Finally, a possible scheme to combine both PT and AV systems is proposed to improve existing PT system.
\section{Working Packages and Tasks}
\subsection{Traffic light scheduling for vehicle-pedestrian mixed-flow networks}
Most traffic signal controllers differentiate vehicles and pedestrians, and focus on their great contributions to vehicle flows, this is reasonable when pedestrian volume is low. However, in the downtown areas where large number of pedestrians interfere with vehicular traffic, optimizing traffic signals only for vehicles may create more conflicts between both traffic participants and potentially reduce the economic interest, since pedestrians in CBD areas are usually potential customers of nearby shopping malls. According to SGS Economics and Planning \cite{SGS}, optimized pedestrian flow can increase \$1.3 billion a year for Melbourne CBD area. On the other hand, rich data can be obtained with the help of the advanced sensor equipments, such as V2X, 5G communication, Lidar and so on. Powerful machine learning algorithms can be effectively utilized to help predict the traffic flows based on the large amount of database, which better serve the optimized signal controller.

\begin{figure*}[!ht]
	\centering
	\includegraphics[width=4.5in]{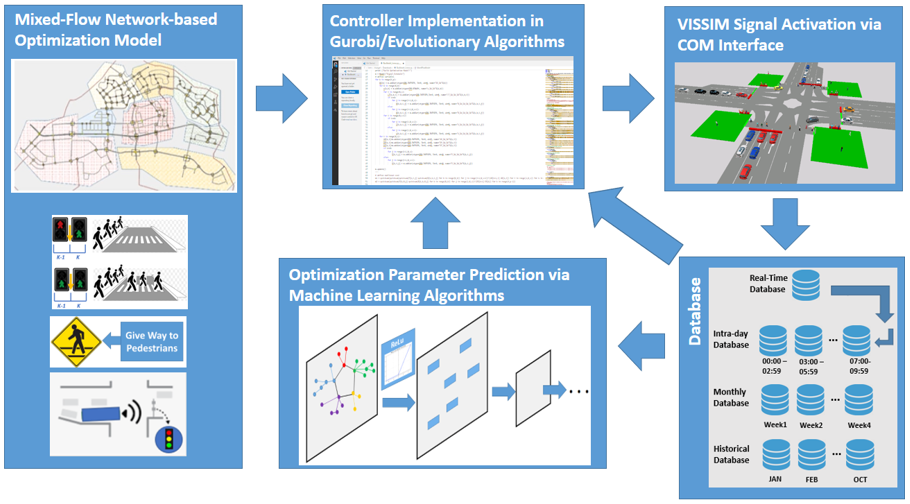}
	\caption{Framework of traffic light scheduling for vehicle-pedestrian mixed-flow networks}
	\label{fig_Task1}
\end{figure*}

In view of this, we propose this signal controller with the aim to fairly minimize both vehicles' and pedestrians' delay \cite{zhang2018traffic}\cite{zhang2019pedestrian}. Fig. \ref{fig_Task1} illustrates the framework of the proposed real-time traffic light scheduling strategy implemented in simulator software VISSIM. The macroscopic flow model for both pedestrians and vehicles are developed, and the impacts of the signal on pedestrian crossing capacity is captured in the model. Benefitted from the advanced sensors, signal priority level could also be incorporated into the model to enable higher priority for public buses. The mixed-flow model is then solved by adopting the commercial optimization solver or evolutionary algorithms. After that, the optimized signal phases and duration are sent to the simulator VISSIM, which mimics the real urban traffic environment. Meanwhile, the traffic information, such as traffic volumes, turning ratios, is stored in the back-end database, which is used to re-train the AI models by adopting the machine learning algorithms to predict the required information. The predicted traffic parameters and current real-time information are all sent to the controller side, which is solved in a rolling horizon manner, and the whole process continues with the evolve of the traffic system to form a closed-loop control strategy.
\subsection{A combined dispatching-operation strategy for public bus management incorporated with the boarding control} \label{sec:task2}
The bus operation systems in most today's studies are proposed based on known bus dispatching time or schedule headways/frequencies. On the other hand, the bus dispatching systems seldom consider operation control for on-road buses. It is understandable that separately deal with these two problems could reduce the computational complexity significantly compared with the complicated combined model. However, the increasing application and perfection of telematics (e.g., Automatic Vehicle Location (AVL), Automatic Passenger Counting (APC), etc.) in the bus management system, the collection of the real-time information becomes possible, also, the constantly upgraded computer with more computing power is keeping breaking the records, thus, it is predictable that to solve a combined optimization problem is implementable \cite{zhang2021multi}. Decision variables, such as bus dispatching time, bus speed between any two adjacent stops, bus dwell time at each stop, OD-based boarding volume, could all be described into a holistic model to enable this combined dispatching-operation bus management system \cite{zhang2021Dynamic}. Also, boarding control has been studied in some literature and proved its high efficiency in improving the bus service quality. However, it  still has not been widely implemented in the real world.

\begin{figure}[!ht]
	\centering
	\includegraphics[width=3in]{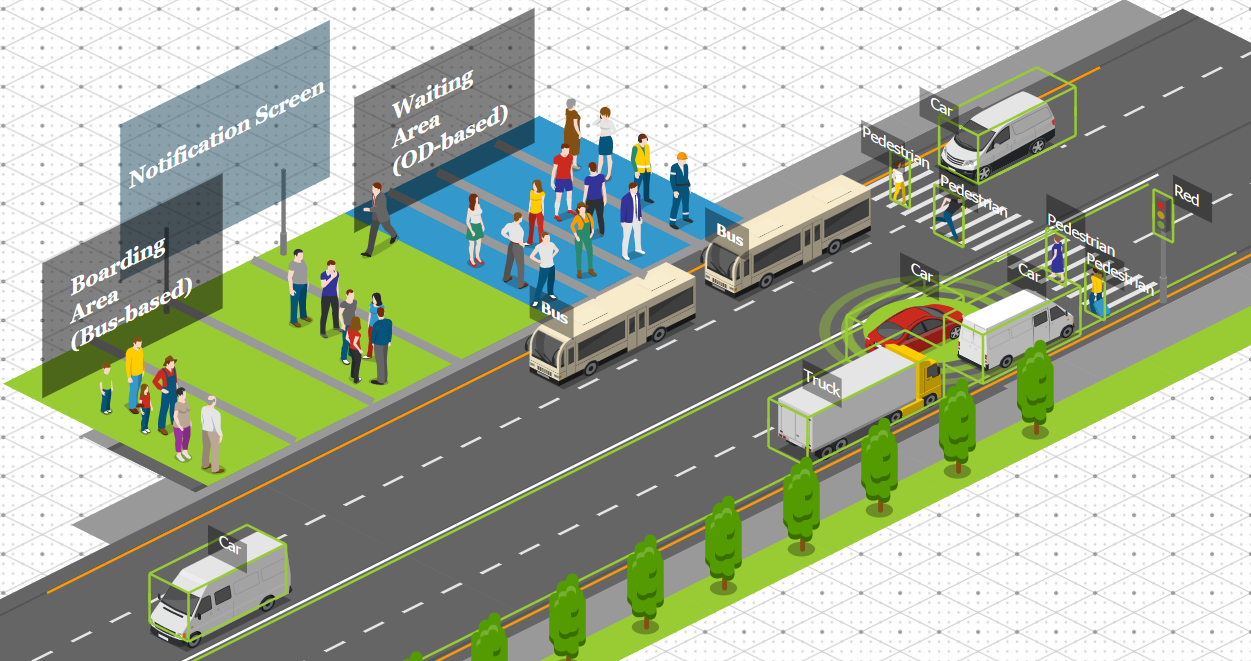}
	\caption{Graphic illustration of future bus stop}
	\label{fig_Task2}
\end{figure}

Fig. \ref{fig_Task2} gives an graph illustration of future bus stop, which makes the boarding control applicable. Imagine a message via mobile apps or screen board located at the bus stop could be sent or displayed to passengers when a bus is approaching the stop, and only passengers at the front queue will be selected to enter the designated boarding area at the bus stop. The identification of the front-queue passenger can be realized by the camera or Lidar installed at the bus stop.
When the bus reaches the bus stop, only passengers who are waiting at the designated area are allowed to board the bus. This requires infrastructural enhancement at the bus stop, not only the area re-design but also the advanced sensors.
\subsection{Autonomous bus fleet management for mobility-on-demand system}
To increase the vehicle utilization and reduce the carbon dioxide emissions, the on-line carsharing services have been carried out in many cities in the past few years. Also, the adoption of the electric vehicles could potentially reduce emissions to promote a sustainable environment, meanwhile, the forth-coming commercialization of AVs has constantly reshaping the transportation system. All of the above emerging concepts and technologies have combined and led to an autonomous mobility-on-demand system (AMOD). AMOD system is not a new concept and has been proposed since 2014 \cite{spieser2014toward}, however, the implementation and research studies about AMOD system mainly focus on private cars and taxi services, and this may bring challenges on the traditional public transport, which has been regarded as the symbol of equity and accessibility. Therefore, it is strongly necessary to identify and explore the synergistic possibilities between AMOD and public transports, especially under this transitional period.

\begin{figure}[!ht]
	\centering
	\includegraphics[width=3in]{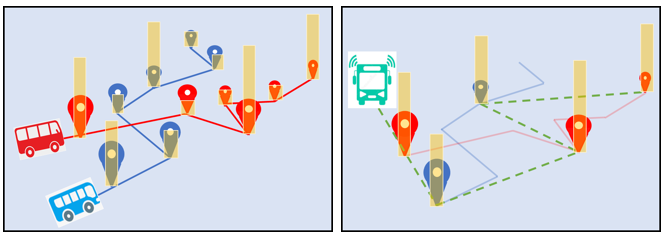}
	\caption{Application of autonomous buses on supporting traditional public bus lines}
	\label{fig_Task3_1}
\end{figure}

\begin{figure}[!ht]
	\centering
	\includegraphics[width=3in]{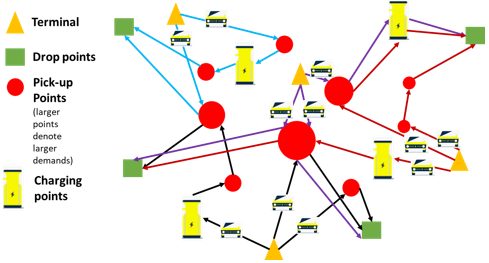}
	\caption{Autonomous bus routing for multi-loading and on-road charging}
	\label{fig_Task3_2}
\end{figure}

Fig. \ref{fig_Task3_1} illustrates one possible case capturing how AVs could support the existing public bus lines. The left subfigure describes two traditional bus lines served by corresponding buses dispatched from the terminal. The orange bar represents the demand size at each bus stop. All control variables mentioned in Section \ref{sec:task2} could be adopted and tuned to facilitate the operation of the traditional bus service. Meanwhile, we could also dispatch AVs to serve bus stops on multiple bus lines but with high demands due to its flexible on-demand characteristics. AVs are normally commercialized as electric cars with limited battery capacity, thus, it is inevitable to develop a smart AV routing strategy catering for the demands of the users as well as the battery capacity level, as illustrated in Fig. \ref{fig_Task3_2}. As AVs do not follow a fixed bus route and on-board passengers need to be sent to their destined stops, thus, on-demand autonomous bus fleet dispatching strategy inevitably requires the boarding control at each bus stop, and this enables the system to consider not only the routing decision but also the volume dynamics, which is different with current one passenger pick-up and drop-off problem. The optimization involving in routing, subsequent bus stop selection and corresponding boarding volume makes the problem more challenging but also interesting.

\section{Conclusion}
In this essay, technological solutions from a pedestrian/transit-oriented perspective have been provided. Firstly, an adaptive traffic
signal control framework is presented, with consideration of a macroscopic mixed-flow optimization model, VISSIM simulation platform, historical database and machine learning-enabled AI prediction model. The experiment results in our paper \cite{zhang2018traffic} have demonstrated that the proposed strategy in a Manhattan-shaped network can strike a good balance between pedestrians' and vehicle drivers' needs. Therefore, we are optimistic to implement this strategy in the real world to enhance the existing traffic signals in the near future. Next, the bus dispatching and its on-road operation, especially the boarding control, are combined together to minimize the passenger delay time and the operating bus vacancy, which makes our strategy more flexible and adaptive to meet the passenger demand. In our previous study \cite{zhang2021Dynamic}, a multi-bus dispatching and boarding control strategy can reduce roughly 50\% of remaining passenger volumes when compared with the timetable-based fixed schedule, which is quite promising and exciting. Moreover, to achieve a synergistic objective between PT system and AV system, a scheme is proposed, where AVs are dispatched to pick up passengers at high-demand stops to support the PT lines. This scheme shall be further studied in my future work to incorporate multiple optimized variables, including AV routing, volume dynamics and boarding limits. Overall, the essay presents three application scenes in the intelligent transportation field. The suggested strategies and guidelines can assist in developing an intelligent future urban transport system in order to provide citizens a smarter, safer and more interactive transportation experience.
\bibliography{ref}
\bibliographystyle{plain}
\begin{IEEEbiography}[{\includegraphics[width=1in,height=1.25in,clip,keepaspectratio]{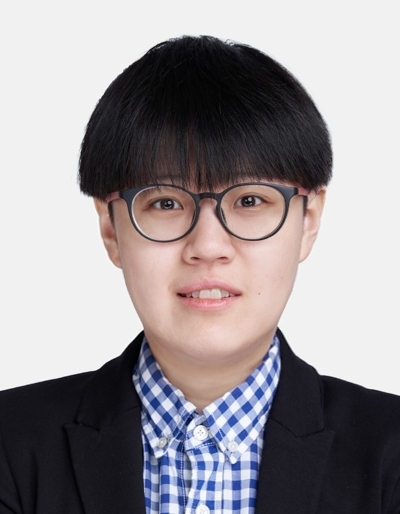}}]{Yi Zhang (S'17-M'21)}
Yi Zhang received her Bachelor degree of Engineering from Shandong University, China in 2014,
and the PhD degree in Electrical and Electronic Engineering from Nanyang Technological University, Singapore in 2020.
She is currently a research scientist at the Institute for Infocomm Research (I2R) in the Agency for Science, Technology and Research, Singapore (A*STAR). Her research interests focus on intelligent transportation system, including urban traffic flow management,
model-based traffic signal scheduling, lane change prediction and bus dispatching and operation management.
\end{IEEEbiography}
\end{document}